\documentclass[prd,onecolumn,showpacs]{revtex4}

\usepackage{graphicx}
\usepackage{array}
\usepackage{bm}
\usepackage[english]{babel}
\usepackage{dcolumn}
\usepackage{amssymb}
\usepackage{amsmath}

\begin{document}

\title{Towers of hybrid mesons}

\author{Claude \surname{Semay}}
%\thanks{F.R.S.-FNRS Senior Research Associate}
\email[E-mail: ]{claude.semay@umh.ac.be}
\author{Fabien \surname{Buisseret}}
%\thanks{F.R.S.-FNRS Postdoctoral Researcher}
\email[E-mail: ]{fabien.buisseret@umh.ac.be}
\affiliation{Groupe de Physique Nucl\'{e}aire Th\'{e}orique,
Universit\'{e} de Mons,
Acad\'{e}mie universitaire Wallonie-Bruxelles,
Place du Parc 20, B-7000 Mons, Belgium}
\author{Bernard \surname{Silvestre-Brac}}
%\thanks{}
\email[E-mail: ]{silvestre@lpsc.in2p3.fr}
\affiliation{LPSC Universit\'e J. Fourier, Grenoble 1, CNRS/IN2P3,
Institut Polytechnique de Grenoble,53 Avenue des Martyrs,
F-38026 Grenoble-Cedex, France}
\date{\today}

\begin{abstract}
A hybrid meson is a quark-antiquark pair in which, contrary to ordinary mesons, the gluon field is in an excited state. In the framework of constituent models, the interaction potential is assumed to be the energy of an excited string. An approximate, but accurate, analytical solution of the Schr\"{o}dinger equation with such a potential is presented. When applied to hybrid charmonia and bottomonia, towers of states are predicted in which the masses are a linear function of a harmonic oscillator band number for the quark-antiquark pair. Such a formula could be a reliable guide for the experimental detection of heavy hybrid mesons.
\end{abstract}


%12.39.Mk Glueball and nonstandard multi-quark/gluon states
%12.39.Pn Potential models
%03.65.Ge Solutions of wave equations: bound states

\maketitle

\section{Introduction}

The gluon being a colored object, Quantum Chromodynamics
(QCD) allows the existence of exotic resonances, such as glueballs or hybrid mesons.
Glueballs are composed of gluons only, while hybrid mesons contain a quark-antiquark pair as well. In particular, the study of heavy hybrid mesons is an active domain 
in theoretical and in experimental particle physics. BELLE and BABAR have already reported the discovery of several intriguing $c\bar c$- or $b\bar b$-like resonances: One can quote the X(3872)~\cite{choi}, but also the Y(4260)~\cite{Aub05} and $\Upsilon(10890)$ resonances~\cite{Che08}, that could be interpreted either as $Q\bar Q$ hybrid mesons or as $Q\bar Q q\bar q$ tetraquarks -- see Refs.~\cite{revexp} for complete reviews. Notice that $Q$ ($q$) denotes a heavy (light) quark. Moreover, one expects that future
experiments like COMPASS, BESIII, GLUEX and PANDA should be very efficient in the detection of heavy hybrid mesons, especially of $c\bar c$-type. 

There are two possible descriptions of hybrid mesons: First, a genuine three-body object made of a quark, an antiquark and a constituent gluon; Second, a two-body object made of a quark and an antiquark in the potential due to the gluon field in an excited state. In the framework of constituent models, it has been shown that these two pictures of the same object are, to a large extent, equivalent \cite{Bui06a,Bui06b,Bui07}. In this paper, we model the heavy hybrid meson as a $Q\bar Q$ pair within an excited gluon field.

In general, the string energy, and therefore the potential energy between the static
quark and antiquark in the excited gluon field is given by \cite{Arv83,All98}
\begin{equation}
\label{eq:potex}
V(r)=\sqrt{a^2 r^2+b},
\end{equation}
where $a$ is the usual string tension while $b=2 \pi a K + C$ is a term exhibiting the string excitation number $K$ and  a constant $C$. These values depend on the model adopted: a pure string theory \cite{Arv83} or a more phenomenological approach \cite{All98,Bui07}. In the present work, we choose the form predicted in Ref.~\cite{Bui07}, $K=2 n_g+l_g$ and $C=3 \pi a$, which is in very good agreement with lattice QCD for the standard value $a=0.2$~GeV$^2$ \cite{Jug98,Jug03}. Finally,
\begin{equation}
\label{eq:bform}
b=2 \pi a (2 n_g+l_g+3/2),
\end{equation}
where $n_g$ and $l_g$ are respectively the radial and orbital quantum numbers of the constituent gluon simulating the excitation of the string. For the study of heavy hybrid mesons, it is therefore very interesting to calculate the eigenenergies of the Schr\"{o}dinger equation governed by the potential (\ref{eq:potex}), or equivalently by the Hamiltonian
\begin{equation}
\label{eq:H}
H=\frac{\bm{p}^2}{2\mu} + \sqrt{a^2 r^2+b},
\end{equation}
where $\mu=m_Q m_{\bar Q}/(m_Q+m_{\bar Q})$ is the reduced mass and where the parameter $b$ is given by Eq.~(\ref{eq:bform}).

The aim of this report is to give an analytical expression for the mass of a heavy hybrid meson in the picture of an excited color field and to derive interesting physical consequences. Our analytical method relies on the auxiliary (or einbein) field method (AFM) which has proved to be very powerful for such kinds of calculations \cite{Sil08a,Sil08b}. The application of the AFM to Hamiltonian~(\ref{eq:H}) is presented in Sec.~\ref{sec:Eigen}, leading to an analytical mass formula. Using this formula, we show in Sec.~\ref{sec:Phenom} that it is possible to predict the general behavior of the heavy hybrid meson masses as a function of the quantum numbers of the system and search for possible towers of states. Our results are summarized in Sec.~\ref{sec:Conc}. The excellent accuracy of the AFM mass formula is discussed in Appendix~\ref{sec:comp} through a comparison with exact numerical results. 

\section{Eigenenergies}
\label{sec:Eigen}

\subsection{Analytical expression}
\label{subsec:analexp}
Assuming that the potential~(\ref{eq:potex}) is the dominant interaction in a heavy hybrid meson, the mass of this system is $M_{\textrm{hm}}=m_Q+m_{\bar Q}+E$, where $E$ is an eigenvalue of the Hamiltonian~(\ref{eq:H}). Using scaling laws (see Ref.~\cite{Sil08b}), dimensionless variables $\epsilon$ and $\beta$ can be defined as
\begin{equation}
\label{eq:eps}
E=\left( \frac{2 a^2}{\mu} \right)^{1/3} \epsilon (\beta), \quad \textrm{with} \quad 
\beta=b \left( \frac{\mu}{2 a^2} \right)^{2/3}.
\end{equation}
$\epsilon (\beta)$ is an eigenvalue of the dimensionless Hamiltonian
\begin{equation}
\label{eq:ph}
h=\frac{\bm{q}^2}{4} + \sqrt{x^2+\beta}.
\end{equation}

Let us follow the general procedure of the AFM \cite{Sil08a} in order to find approximate expressions
for the eigenvalues of the Hamiltonian~(\ref{eq:ph}). We first choose an auxiliary function $P(x)=x^2$; the auxiliary field $\nu$ is then defined by
\begin{equation}
\label{eq:funcK}
\nu = K(x) = \frac{V'(x)}{P'(x)} = \frac{1}{2 \sqrt{x^2+\beta}}.
\end{equation}
For the moment $\nu$ is an operator, and Eq.~(\ref{eq:funcK}) can be inverted to give
$x$ as a function of $\nu$: $x = I(\nu)$. Explicitly
\begin{equation}
\label{eq:funcI}
I(\nu)=\sqrt{\frac{1}{4 \nu^2}-\beta}.
\end{equation}

The AFM needs the definition of a Hamiltonian $\tilde{h}(\nu) = \bm{q}^2/4 + \nu P(x) +
V(I(\nu))-\nu P(I(\nu))$. In our particular case,
\begin{equation}
\label{eq:tildeH}
\tilde{h}(\nu) = \frac{\bm{q}^2}{4} + \nu x^2 + \frac{1}{4 \nu} +\beta \nu.
\end{equation}
If we choose the auxiliary field in order to extremize $\tilde{h}(\nu)$, $\delta \tilde{h}/
\delta \nu |_{\nu = \hat{\nu}} = 0$, then the value of this Hamiltonian for such an
extremum is precisely the original Hamiltonian: $\tilde{h}(\hat{\nu}) = h$. Instead of
considering the auxiliary field as an operator, let us consider it as a real number.
In this case, the eigenenergies of $\tilde{h}$ are exactly known for all $(n,l)$ quantum
numbers:
\begin{equation}
\label{eq:eigenener}
\epsilon(\nu)=\sqrt{N^2 \nu}+ \frac{1}{4 \nu} + \beta \nu,
\end{equation}
where, as usual, $N=2n+l+3/2$ is the principal quantum number of the state.

The philosophy of the AFM is very similar to a mean field  procedure. We first seek the value
$\nu_0$ of the auxiliary field which minimizes the energy, $\partial \epsilon(\nu)/\partial \nu
|_{\nu=\nu_0}$, and consider that the value $\epsilon(\nu_0)$ is a good approximation
of the exact eigenvalue \cite{Sil08a,Sil08b}.
At this stage it is useful to define the new variable
\begin{equation}
\label{eq:defx0}
x_0 = N^{-1/3} \nu_0^{-1/2}
\end{equation}
and the parameter
\begin{equation}
\label{eq:defY}
Y=\frac{16\beta}{3N^{4/3}}.
\end{equation}
The minimization condition is now concerned with $x_0$ and
results from the fourth order reduced equation
\begin{equation}
\label{eq:redeq}
4 x_0^4 - 8 x_0 - 3Y = 0.
\end{equation}
The solution of this equation can be obtained by standard algebraic techniques.
It looks like
\begin{equation}
\label{eq:valx0}
x_0 = G(Y) =  \frac{1}{2} \sqrt{V(Y)} + \frac{1}{2} \sqrt{ 4 (V(Y))^{-1/2}
- V(Y)},
\end{equation}
with
\begin{equation}
\label{eq:defVY}
V(Y)=\left(2 + \sqrt{4 + Y^3} \right)^{1/3} - Y \left(2 + \sqrt{4 + Y^3}
\right)^{-1/3}.
\end{equation}
Substituting this value into the expression of $E(\nu_0)$ leads to the analytical
form of the searched eigenenergies, namely
\begin{eqnarray}
\label{eq:Enu0}
\epsilon_{{\rm AF}}=\epsilon(\nu_0)&=&2\sqrt{\frac{\beta}{3 Y}} \left[G^2(Y)+\frac{1}{G(Y)} \right].
\end{eqnarray}
The problem is entirely solved.

As it is shown in Ref.~\cite{Sil08b}, the same formula would be obtained for the choice 
$P(r)=\textrm{sgn} (\lambda)r^\lambda$ with $\lambda > -2$, but with different 
forms for the quantity $N$. With the choice $\lambda = 2$ made above, $N=2 n+l+3/2$. 
In this case, using results from Ref.~\cite{Bui08}, it can be shown that formula~(\ref{eq:Enu0}) 
gives an upper bound of the exact result. For $\lambda = -1$, $N=n+l+1$ and 
the formula gives a lower bound. The qualities of these bounds are examined in Appendix~\ref{sec:comp}. It is also shown in this appendix that the expression
\begin{equation}
\label{eq:exprN}
N=A(\beta) n + l + C(\beta), 
\end{equation}
with $A(\beta)$ and $C(\beta)$ defined by Eq.~(\ref{eq:coefacd}), leads to an analytical formula which reproduces very accurately (up to 1\%) the exact results.

%An approximate simpler form of Eqs.~(\ref{eq:Enu0}) avoiding the complicated $G$ function and  
%giving the lowest order exact results in both limits 
%$Y\to 0$ and $Y\to \infty$ for finite value of $b$, is given by 
%\begin{equation}
%\label{eq:Enu0alt}
%E(\nu_0)\approx \sqrt{\frac{b}{3Y}} \left( \sqrt{3 Y+(3\times2^{2/3}-\eta)^2}+\eta \right),
%\end{equation}
%where $\eta$ is an arbitrary parameter.
%A very good approximation is obtained for $\eta$ around 1: for a fixed value of $b$, the relative error 
%between Eqs.~(\ref{eq:Enu0}) and Eq.~(\ref{eq:Enu0alt}) is below 2\%.

\subsection{Asymptotic expansions}
\label{subsec:asymp}
Equation~(\ref{eq:Enu0}) is complicated but quite accurate. In order to get a better insight into this formula, it is interesting to compute several limits:
\begin{align}
\label{eq:lYg}
\lim_{Y\gg 1} \epsilon_{\textrm{AF}} &= \sqrt{\beta} +\frac{N}{\sqrt{2}\beta^{1/4}}-\frac{N^2}{16\beta}, \\
\label{eq:lYp}
\lim_{Y\ll 1} \epsilon_{\textrm{AF}} &= \frac{3 N^{2/3}}{2^{4/3}} + \frac{\beta}{2^{2/3}N^{2/3}}-\frac{\beta^2}{3N^2}.
\end{align}
The first two terms in the r.h.s. of Eq.~(\ref{eq:lYg}) are the solution of a harmonic potential, while Eq.~(\ref{eq:lYp})
with $\beta=0$ corresponds to the solution of a linear interaction obtained with the AFM~\cite{Sil08a}.

For the lowest excited state of the gluon field ($K=0$) and for physical values of the parameters ($a\approx 0.15$-$0.20$~GeV$^2$, $m_c\approx 1.1$~GeV, $m_b\approx 4.5$~GeV), it comes that $\beta\approx 7$-$20$. In this case, $\beta$ is large enough for the choice $N=2 n+l+3/2$ to be relevant (see Eq.~(\ref{eq:coefacd})). A harmonic oscillator band number $B=2 n+l$ can thus be introduced to label the states. We can further assume that hybrid mesons with $K=0$ and $B \gtrsim 4$ cannot be easily produced and discriminated from the low-lying $K=1$ hybrids which would lie in the same mass range. So in the following, we will only consider that $\beta \in [7,20]$, $B \le 4$, and $K=0$. It is indeed more probable that hybrids with the lowest possible excitation of the gluon field will be first observed. 

As we are interested in the study of towers of states, it could be expected that the power expansion~(\ref{eq:lYp}), valid for large $N$, should be used. But, $\beta$ is so large for heavy quarks that it is actually not the case. For example, when $\beta=20$, one has $Y\approx 5\gg 1$ even with the large value $N=10$. By comparing with accurate numerical solutions obtained for the Hamiltonian $h$ with the Lagrange mesh method \cite{sem01}, it has been checked that the power expansion~(\ref{eq:lYg}) is far better. Moreover, the ratio of the third term over the second one in this expansion is at most 10\%. So with very good approximation the eigenvalues of $h$ are given by %(see Fig.~\ref{fig1})
\begin{equation}
\label{eq:epsapp}
\epsilon_{\textrm{app}} = \sqrt{\beta} +\frac{B+3/2}{\sqrt{2}\beta^{1/4}}.
\end{equation}
The approximate formula~(\ref{eq:epsapp}) only depends on $B$. This exact degeneracy is actually broken by the non-harmonic character of the Hamiltonian $h$, but the breaking is small: %(see Fig.~\ref{fig1}). 
It can be numerically checked that the maximal relative error of Eq.~(\ref{eq:epsapp}) with respect to the exact values of $\epsilon-\sqrt{\beta}$ is around 10\%. 
So, $B$ is a relevant classification number.
%\begin{figure}[t]
%\begin{center}
%\includegraphics*[width=8.0cm]{fig1.eps}
%\end{center}
%\caption{Eigenvalue of the dimensionless Hamiltonian $h$ for $\beta=18$ as a function of the band number $B=2 n+l=N-3/2$. Dots: $\epsilon$ exact (the non-degeneracy is hardly distinguishable at this scale); solid line: $\epsilon_{\textrm{AF}}$ [Eq.~(\ref{eq:epsform})]; dashed line: $\epsilon_{\textrm{app}}$ [Eq.~(\ref{eq:epsapp})].}
%\label{fig1}
%\end{figure}

\subsection{Final mass formula}
The Coulomb interaction must also play a role in heavy hybrids, since heavy quarks are expected to orbit close to each other and must feel strongly this short range interaction. The Hamiltonian~(\ref{eq:ph}) might then be completed with the potential $\kappa/x$ where
\begin{equation}
\label{eq:kappa}
\kappa = \frac{\alpha_S}{6}\left( \frac{\mu^2}{4 a} \right)^{1/3},
\end{equation}
$\alpha_S$ being the strong coupling constant and the $1/6$ factor coming from color Casimir operator. Since $\alpha_S \lesssim 0.2$ in the heavy meson sector \cite{Bui06b}, it appears that $\kappa\lesssim 0.02$-$0.07$. For the large values of $\beta$ considered here, the contribution of the potential $\kappa/x$ can be computed in perturbation. Using the AFM results, one obtains
\begin{equation}
\label{eq:meankap}
\left\langle \frac{\kappa}{x} \right\rangle \approx \frac{2^{1/4}\kappa}{\beta^{1/8} \sqrt{B+3/2}}.
\end{equation}
This contribution is at most around 4\% of $\epsilon-\sqrt{\beta}$ for $c\bar c$ hybrids and 10\% for $b\bar b$ hybrids. So, in first approximation, it is relevant to neglect the Coulomb interaction in the search for towers of hybrid mesons.

Under these conditions, the mass of a hybrid meson is given by
\begin{eqnarray}
\label{eq:mqqg}
M_{\textrm{hm}} &\approx& m_Q+m_{\bar Q}+\sqrt{2\pi a (K+3/2)} \nonumber \\
&&+\sqrt{\frac{a^{3/2}}{\mu\sqrt{2\pi (K+3/2)}}}\, (B+3/2).
\end{eqnarray}
For high excitation of the gluon field ($K\gg 0$), $\beta$ increases and the accuracy of the approximation~(\ref{eq:mqqg}) improves. Using the AFM, an accurate mass formula for a ordinary meson with just the linear confinement is given by \cite{Sil08a}
\begin{equation}
\label{eq:mqq}
M_{\textrm{om}}\approx m_Q+m_{\bar Q}+ \frac{3}{2}\left( \frac{a^2}{\mu} \right)^{1/3} 
\left( \frac{\pi}{\sqrt{3}}n +l+ \frac{\sqrt{3}\pi}{4} \right)^{2/3}.
\end{equation}
Notice that $K=0$ is not the ground state of the flux tube, but rather its first excited level. The ground state simply corresponds to $b=0$ in Eq.~(\ref{eq:potex}), that is a linear confining potential, and leads to formula (\ref{eq:mqq}).

\section{Phenomenology}
\label{sec:Phenom}

Some comments have to be done at this stage. Hamiltonian~(\ref{eq:H}) describes a genuine heavy hybrid meson (no mixing with other hadronic states) within a spin-independent formalism. It will thus lead to qualitative global predictions rather than to a detailed mass spectrum. All spin effects are neglected but they should be weak for heavy hybrids since they are proportional to $1/(m_Q m_{\bar Q})$ -- see for example Ref.~\cite{kalash} for a numerical check of that point. We think however that such global predictions are quite robust precisely because they do not depend on a fine-tuned model. Nevertheless, it is important to wonder whether our method will preferentially apply to some $J^{PC}$ quantum numbers or not. Non-exotic quantum numbers, like $1^{--}$ for example, must be examined very carefully because such quantum numbers allow for a possibly strong mixing with ordinary mesons. %is expected in this case~\cite{tetra}. 
Such a mixing is by definition absent for exotic quantum numbers like $1^{-+}$, $2^{+-}$, $3^{-+}$, \dots, although mixing with tetraquarks (or even glueballs) cannot be excluded. To our knowledge, the mixing between hybrid mesons and tetraquarks is far from being well-known theoretically, and we will not discuss it in the present work.   
\begin{figure}[t]
\begin{center}
\includegraphics*[width=8.5cm]{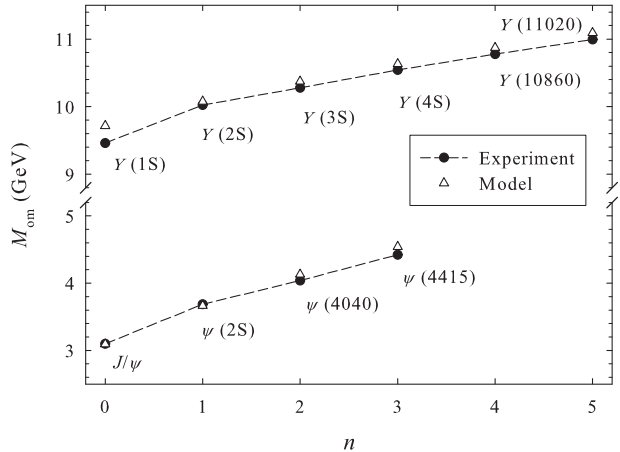}
\end{center}
\caption{Comparison between experimental $c\bar c$ and $b\bar b$ radial trajectories (full circles) and the predictions of Eq.~(\ref{eq:mqq}) for $a=0.2$~GeV$^2$, $m_c=1.152$~GeV, and $m_b=4.620$~GeV. Experimental data come from the PDG~\cite{PDG} and the quantum numbers assignment of Ref.~\cite{lak} for the $c\bar c$ mesons is followed. Dashed lines are plotted to guide the eyes.}
\label{fig1}
\end{figure}   

Let us first estimate the value of the our model's parameters. The spin effects in mesons are minimal in the $1^{--}$ channel; it is thus relevant to fit the parameters by requiring Eq.~(\ref{eq:mqq}) to reproduce the radial trajectories of the $1^{--}$ $c\bar c$ and $b\bar b$ mesons. We first take $a=0.2$~GeV$^2$ so that potential~(\ref{eq:potex}) optimally fits the corresponding lattice QCD data~\cite{Bui06b}. Then, it is readily seen in Fig.~\ref{fig1} that the experimental data are well described by setting $m_c=1.152$~GeV and $m_b=4.620$~GeV -- the $\Upsilon(1S)$ is poorly reproduced because of the neglect of the Coulomb term. Such values are not unrealistic when compared to the PDG values of $1.27^{+ 0.07}_{- 0.11}$~GeV and $4.20^{+ 0.17}_{- 0.07}$~GeV~\cite{PDG}, and quite common in effective approaches. 

As already said in the introduction, the status of the $1^{--}$ resonance $Y(4260)$ is not clear yet. If it is an ordinary meson, it should have $S=1$, $n=3$ and $l=0$ \cite{Lla05}, and Eq.~(\ref{eq:mqq}) then leads to a mass of 4.512~GeV, quite far from the experimental value. If it is a hybrid meson, the quantum numbers must be $S=1$, $n=0$ and $l=1$ \cite{Gen07}, and Eq.~(\ref{eq:mqqg}) leads to a more similar mass of 4.239~GeV. But, the repulsive Coulomb term would slightly decrease this agreement. The ordinary meson interpretation for the $1^{--}$ resonance $\Upsilon(10890)$ has been discarded in Ref.~\cite{Che08}; Eq.~(\ref{eq:mqqg}) is compatible with the hybrid meson picture since it leads to a mass of 10.894~GeV for the corresponding state. Our model does not disagree with a hybrid meson assignment for both states, but no definitive conclusion can be drawn at this stage, mainly because of the neglect of the mixing with ordinary mesons. 

We now propose a \textit{modus operandi} to separate the hybrid mesons from the ordinary ones. Let us first assume that we have a sufficient amount of experimentally found $Q\bar Q$-like resonances -- this should be possible in a near future thanks to the forthcoming experiments we mentioned in the introduction. Then, for each $J^{PC}$ state, one has to find the lowest possible value of $B$ that is compatible with these quantum numbers in a hybrid meson picture. We recall that $P=(-1)^{l+l_g}$ and $C=(-1)^{l+S_{q\bar q}+1}$ in this case, and that $l_g=0$ when $K=0$. Once this step is achieved, one can plot the experimental masses versus $B$. Our prediction is that the hybrid mesons, or at least states that are dominated by a hybrid meson component, will be located along a straight line while other states like ordinary mesons or tetraquarks will not (typically for example, one has $M_{\textrm{om}}-m_Q-m_{\bar Q}\propto n^{2/3}$ and $l^{2/3}$). 

\begin{figure}[t]
\begin{center}
\includegraphics*[width=8.5cm]{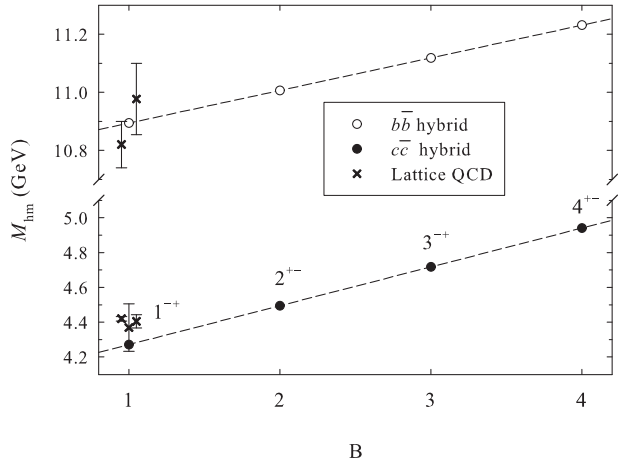}
\end{center}
\caption{Masses of exotic $Q\bar Q$ hybrids versus the band number $B$ as predicted by formula~(\ref{bmass}) (circles and dashed lines). Parameters of Fig.~\ref{fig1} are used. An explicit example of exotic $J^{PC}$ trajectory is given; the lattice data of Refs.~\cite{clat0,clat1,clat2,blat1,blat2} are also plotted for comparison (crosses).}
\label{fig2}
\end{figure}

Formula~(\ref{eq:mqqg}) can be rewritten as follows for $K=0$
\begin{equation}\label{bmass}
M_{{\rm hm}}(B) = \sqrt{\frac{a^{3/2}}{\mu\sqrt{3\pi }}}\ (B-B_0)+M_{{\rm hm}}(B_0).
\end{equation}
It is plausible that an exotic $1^{-+}$ hybrid meson, thus with $B=1$, will be first discovered. Effective approaches as well as lattice QCD indeed reach the conclusion that the $1^{-+}$ state is the lightest exotic $Q\bar Q$ hybrid meson (see \textit{e.g.} Refs.~\cite{revlat,kalash}). Then, by setting $B_0=1$ and $M_{{\rm hm}}(B_0)=M_{1^{-+}}$, Eq.~(\ref{bmass}) states that the first exotic states will be located on a straight line, as schematically represented in Fig.~\ref{fig2}. Note that in our approach, the $1^{-+}$ $c\bar c$ state lies below the experimental estimation of the $DD^*$ threshold, which is of about 4.3 GeV~\cite{PDG} ($D^*$ denotes the P-wave excitation of the $D$ meson). 

The results of formula~(\ref{bmass}) can also be compared with lattice QCD. Most of the efforts in lattice QCD were devoted to the computation of the lowest-lying $1^{-+}$ hybrid meson masses. We can quote: 4.420$\pm$0.013~GeV~\cite{clat0}, $4.369 \pm0.136$~GeV~\cite{clat1}, and $4.405\pm0.038$~GeV~\cite{clat2} for the $1^{-+}$ $c\bar c$ states, and $10.82\pm0.08$~GeV~\cite{blat1} and $10.977\pm0.123$~GeV~\cite{blat2} for the $1^{-+}$ $b\bar b$ states. It is worth saying that the $1^{-+}$ and $1^{--}$ $b\bar b$ hybrid mesons are degenerate in Ref.~\cite{blat1}, showing the weakness of spin effects for systems of bottom quarks. Formula~(\ref{bmass}) is in agreement with the current estimations as shown in Fig.~\ref{fig2}; it would be very interesting that other masses become available in lattice QCD in order to check whether they are located on the straight lines we predict or not. 

\section{Conclusion}
\label{sec:Conc}

Starting from a potential model of hybrid mesons which is in agreement with lattice QCD, we predict that the mass of the lowest-lying genuine heavy hybrid mesons is a linear function of the band number $B=2 n+l$, $n$ and $l$ being the quantum numbers of the quark-antiquark pair. The slope and the intercept at the origin depend on the excitation quantum number $K$ of the gluon field. These heavy hybrid mesons form towers of states organized as lines in plots where the masses are presented versus the harmonic oscillator band number. We believe that this property can be an interesting tool to disentangle resonances which are dominated by a heavy hybrid meson component from other hadronic states in future experiments, just as the existence of Regge trajectories is an important guide to identify light mesons. Apart from experiment, an important check of the present results would be the computation of higher-spin exotic hybrid mesons on the lattice. We hope that such results will be available in the future. 

\acknowledgments
CS and FB thank the F.R.S.-FNRS for financial support.

\begin{appendix}
\section{Comparison to exact results}
\label{sec:comp}

The AFM cannot give strong constraints on the dependence
of $N$ in terms of $(n,l)$. In particular, had we chosen $P(r)=\textrm{sgn}(\lambda)\,
r^\lambda$, the better choice for $N$ would have been $N = A(\lambda)n + l +
C(\lambda)$, with the quantities $A(\lambda)$ and $C(\lambda)$ given in
Ref.~\cite{Sil08a}. The square root potential $\sqrt{x^2+\beta}$ ensures a smooth transition from a
linear form ($\lambda=1$ but in this case we have only approximate expressions) to
a quadratic form ($\lambda=2$ and in this case the values are exact) as $\beta$
increases from $0$ to $\infty$.

\begin{figure}[ht]
\begin{center}
\includegraphics*[width=6.4cm]{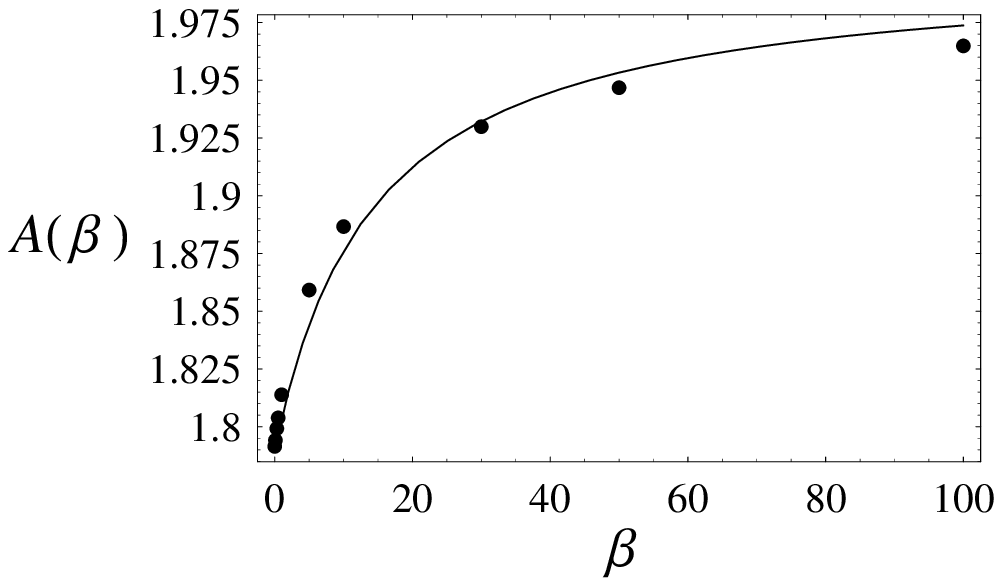}
\includegraphics*[width=6.4cm]{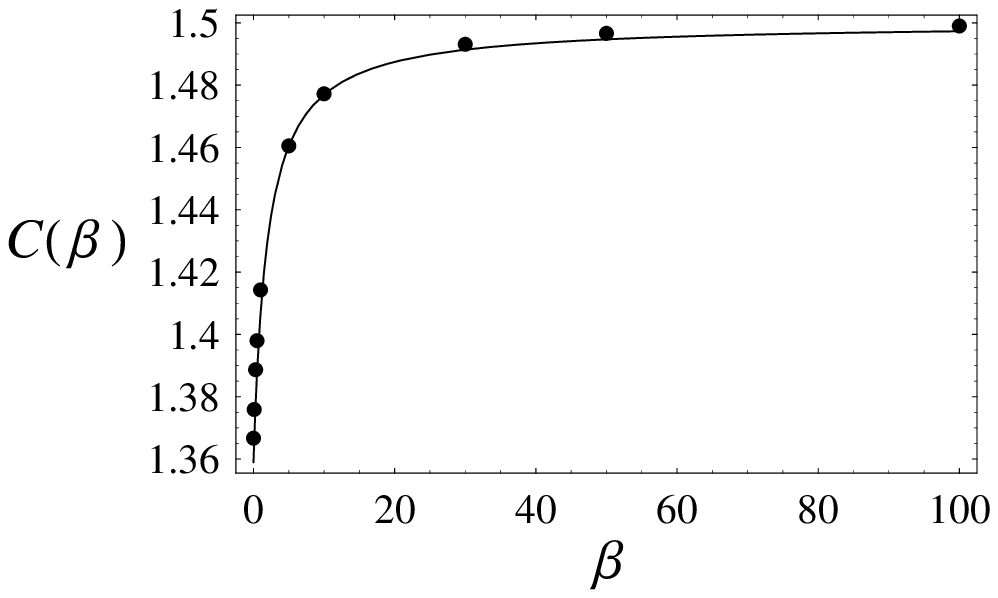}
\caption{\label{fig:exp} Best values of the coefficients $A(\beta)$ and $C(\beta)$
to parameterize the eigenvalues of Hamiltonian~(\ref{eq:ph}): 
numerical fit with Eq.~(\ref{eq:chis}) (dots); functions~(\ref{eq:coefacd}) (solid line).} 
\end{center}
\end{figure}

The procedure we adopt is based on the following points:
\begin{itemize}
  \item We calculate the exact values $\epsilon_{nl}^*(\beta)$ for $0 \leq n \leq
n_{\max}$, $0 \leq l \leq l_{\max}$ and for a given set of $\beta$ values. This
program is achieved using a very powerful method known as the Lagrange mesh
method (described in detail in Ref.~\cite{sem01}). For our purpose, we consider
that $n_{\max}=l_{\max}=4$ is a good choice. For any calculated value, we have
an accuracy better than $10^{-5}$.
  \item We calculate the approximate values $\epsilon_{nl}(\beta)$ using Eqs.~(\ref{eq:Enu0})
with $Y$ given by Eq.~(\ref{eq:defY}) in which $N$ is deduced from Eq.~(\ref{eq:exprN}),
for the same set of $\beta$ values. Building the $\chi$-square
\begin{equation}
\label{eq:chis}
\chi(\beta) = \frac{1}{(n_{\max}+1)(l_{\max}+1)}\sum_{n=0}^{n_{\max}} \sum_{l=0}^{l_{\max}}
\left(\epsilon_{nl}^*(\beta)-\epsilon_{nl}(\beta)\right)^2,
\end{equation}
we request the coefficients $A(\beta)$ and $C(\beta)$ of $N$ to minimize this function. 
The obtained values are represented by black dots in Figs.~\ref{fig:exp}.
  \item In order to obtain functions which are as simple as possible, continuous in $\beta$,
and which reproduce at best the above calculated values, we choose hyperbolic forms
and require a best fit on the set of the sample. Explicitly, we find
\begin{equation}
\label{eq:coefacd}
A(\beta) = \frac{8 \beta + 102}{4 \beta + 57}, \quad
C(\beta) = \frac{30 \beta + 53}{20 \beta + 39}.
\end{equation}
These integers are rounded numbers whose magnitude is chosen in order to not exceed too much 100.
The corresponding values are plotted as continuous curves in Figs.~\ref{fig:exp}.
They have been constrained to exhibit the right behavior $A \to 2$ and
$C \to 3/2$ for very large values of $\beta$. Formulas~(\ref{eq:coefacd}) give $A(0)=102/57\approx 1.789$ and
$C(0)=53/39\approx 1.359$. These values are such that 
$A(0)\approx \pi/\sqrt{3}\approx 1.814$ and $C(0)\approx \sqrt{3}\pi/4\approx 1.360$, as expected from the results of Ref.~\cite{Sil08a} in the case of a nonrelativistic linear potential. 
\end{itemize}

\begin{table}[ht]
\caption{\label{tab:comp} Comparison between the exact values $\epsilon_{nl}^*(\beta)$ (2nd line)
and analytical approximate expressions $\epsilon_{nl}(\beta)$ for the
eigenvalues of Hamiltonian~(\ref{eq:ph}) with $\beta=1$. For each set $(n,l)$, the exact result
is obtained by numerical integration. 3rd line: approximate results
are given by Eqs.~(\ref{eq:Enu0}) with Eqs.~(\ref{eq:defY}), (\ref{eq:exprN}) and
(\ref{eq:coefacd}); 1st line: upper bounds obtained with $N=2 n+l+3/2$; 4th line:
lower bounds obtained with $N=n+l+1$.}
\begin{ruledtabular}
\begin{tabular}{cccccc}
    & $l=0$ & $l=1$ & $l=2$ & $l=3$ & $l=4$\\
\hline
$n=0$ & 1.94926 & 2.49495 & 2.99541 & 3.46197 & 3.90193 \\
      & 1.91247 & 2.45074 & 2.94841 & 3.41419 & 3.85430 \\
      & 1.89549 & 2.44621 & 2.95032 & 3.41969 & 3.86189 \\
      & 1.65395 & 2.22870 & 2.75000 & 3.23240 & 3.68492 \\

$n=1$ & 2.99541 & 3.46197 & 3.90193 & 4.32027 & 4.72059 \\
      & 2.89556 & 3.34652 & 3.77899 & 4.19405 & 4.59335 \\
      & 2.85420 & 3.32970 & 3.77678 & 4.20097 & 4.60620 \\
      & 2.22870 & 2.75000 & 3.23240 & 3.68492 & 4.11355 \\

$n=2$ & 3.90193 & 4.32027 & 4.72059 & 5.10556 & 5.47723 \\
      & 3.74112 & 4.14232 & 4.53310 & 4.91307 & 5.28251 \\
      & 3.69078 & 4.11913 & 4.52783 & 4.91998 & 5.29790 \\
      & 2.75000 & 3.23240 & 3.68492 & 4.11355 & 4.52250 \\

$n=3$ & 4.72059 & 5.10556 & 5.47723 & 5.83725 & 6.18692 \\
      & 4.50374 & 4.87138 & 5.23246 & 5.58628 & 5.93264 \\
      & 4.44883 & 4.84403 & 5.22459 & 5.59242 & 5.94903 \\
      & 3.23240 & 3.68492 & 4.11355 & 4.52250 & 4.91485 \\

$n=4$ & 5.47723 & 5.83725 & 6.18692 & 6.52732 & 6.85935 \\
      & 5.20859 & 5.55148 & 5.88996 & 6.22329 & 6.55111 \\
      & 5.15078 & 5.52098 & 5.87970 & 6.22821 & 6.56756 \\
      & 3.68492 & 4.11355 & 4.52250 & 4.91485 & 5.29295 \\
\end{tabular}
\end{ruledtabular}
\end{table}

%\begin{figure}[ht]
%\begin{center}
%\includegraphics*[width=8cm]{fig2.eps}
%\caption{\label{fig:comp} Spectra $\epsilon_{nl}$ of Hamiltonian~(\ref{eq:Hbeta}) with $\beta=1$.
%For each value of $n$, the eigenvalues are presented for $l$ varying from 0 to 4. 
%Diamonds: exact results obtained by numerical integration.
%Circles: approximate results given by Eqs.~(\ref{eq:Enu0}) with Eqs.~(\ref{eq:simplY}), (\ref{eq:exprN}) and
%(\ref{eq:coefacd}). The error bars extend from lower bounds obtained with $N=n+l+1$ to 
%upper bounds obtained with $N=2 n+l+3/2$.} 
%\end{center}
%\end{figure}

Since our results are exact for $\beta \to \infty$, one has obviously $\chi=0$
in this limit. The error is maximal for small values of $\beta$ but, over the
whole range of $\beta$ values, the results given by 
our analytical expression can be considered as excellent.
Just to exhibit a quantitative comparison, we report in Table~\ref{tab:comp} the exact $\epsilon_{nl}^*(\beta)$ and approximate $\epsilon_{nl}(\beta)$ values
obtained for $\beta=1$, a value for which the corresponding potential is neither
well approximated by a linear one nor a harmonic one. 
As can be seen, our approximate expressions are better than $1\%$ for any
value of $n$ and $l$ quantum numbers. Such a good description is
general and valid whatever the parameter $\beta$ chosen.

The upper bounds obtained with $P(r)=r^2$ are far better than the lower bounds computed with $P(r)=-1/r$.
This is expected since the potential $\sqrt{a^2 r^2+b}$ is closer to a harmonic interaction than to a Coulomb one.
Better lower bounds could be obtained with $P(r)=r$. But, 
the exact form of $N$ is not known for this potential, except for $l=0$ for which $N$ can be 
expressed in term of zeros of the Airy function. With the approximate form 
$N=(\pi/\sqrt{3}) n+l+\sqrt{3}\pi/4$ \cite{Sil08a,Sil08b}, we have checked that results obtained are good but 
the variational character cannot be guaranteed.

\end{appendix}

\end{document}